\documentclass[symmetry,article,submit,moreauthors,pdftex,10pt,a4paper]{mdpi} 
\usepackage{graphicx}
\usepackage{amssymb}
\preto{\abstractkeywords}{\nolinenumbers}

\firstpage{1} 
\articlenumber{x}
\doinum{10.3390/------}
\pubvolume{xx}
\pubyear{2021}
\copyrightyear{2021}
\externaleditor{Academic Editor: Quentin G.\ Bailey}
\history{Received: date; Accepted: date; Published: date}

\def\al{\alpha}
\def\be{\beta}
\def\ga{\gamma}
\def\de{\delta}
\def\ep{\epsilon}
\def\ze{\zeta}
\def\et{\eta}
\def\th{\theta}
\def\ka{\kappa}
\def\la{\lambda}

\def\ta{\tau}

\def\Ga{\Gamma}

\def\Gat{{\tilde \Ga}}

\def\cG{{\cal G}}
\def\cL{{\cal L}}
\def\cM{{\cal M}}

\def\mn{{\mu\nu}}
\def\ab{{\al\be}}
\def\gd{{\ga\de}}
\def\bg{{\be\ga}}
\def\abgd{{\al\be\ga\de}}

\def\abg{{\al\be\ga}}
\def\bgd{{\be\ga\de}}

\def\prt{\partial}

\def\pt#1{\phantom{#1}}

\newcommand{\beq}{\begin{equation}}
\newcommand{\eeq}{\end{equation}}
\newcommand{\bea}{\begin{eqnarray}}
\newcommand{\eea}{\end{eqnarray}}
\newcommand{\rf}[1]{(\ref{#1})}

\def\tb{\overline{t}}
\def\sb{\overline{s}}
\def\st{\tilde{s}}
\def\ub{\overline{u}}
\def\tt{\tilde{t}}
\def\ut{\tilde{u}}

\def\Gt{\tilde{G}}

\def\etal {{\it et al.}}

\def\fr#1#2{{{#1} \over {#2}}}

\def\frac#1#2{{\textstyle{{#1}\over {#2}}}}

\def\rf#1{(\ref{#1})}

\Title{Construction of higher-order metric fluctuation terms in spacetime symmetry-breaking effective field theory}

\Author{Quentin G.\ Bailey $^{1}$}

\AuthorNames{Quentin G.\ Bailey}

\address{%
$^{1}$ \quad Embry-Riddle Aeronautical University, Prescott, AZ; baileyq@erau.edu} 

\corres{Correspondence: baileyq@erau.edu; Tel.: 928-777-3932}

\abstract{We examine the basic conservation laws for diffeomorphism symmetry in the context of spontaneous diffeomorphism and local Lorentz-symmetry breaking.  
The conservation laws are used as constraints on a generic series of terms 
in an expansion around a flat background.  
We find all such terms for a two-tensor coupling 
to cubic order in the metric and tensor field fluctuations.  
The results are presented in a form 
that can be used for phenomenological calculations.  
One key result is that if we preserve the underlying diffeomorphism symmetry 
in a spontaneous-symmetry breaking scenario, 
one cannot decouple the two-tensor fluctuations from the metric fluctuations
at the level of the action, 
except in special cases of the quadratic actions.}


\begin{document}

\section{Introduction}

The diffeomorphism, 
the operation of mapping points on a spacetime manifold, 
is a fundamental symmetry transformation in gravitational theories of physics.  
Together with local Lorentz transformations in the tangent space at each point, 
these comprise the gauge symmetry groups that allow 
General Relativity to be formulated as a local gauge theory of gravitation
\cite{Hehl:1976}.
Since local gauge symmetries can be broken in quantum field theories, 
for instance via spontaneous symmetry breaking, 
it is natural ask if the same could occur for local Lorentz 
and diffeomorphism symmetries \cite{Kostelecky:1988}.
Much recent work has been devoted to the possibility 
of spacetime-symmetry breaking, 
and its possible detection in sensitive tests \cite{Hees:2016, Kostelecky:2008, Tasson:2016, Will:2014,Yunes:2016}.
Among the theoretical mechanisms for this symmetry breaking 
is the elegant spontaneous symmetry breaking one, 
in which the underlying action maintains the local symmetries, 
but due to the presence of a potential energy function with a nonzero
minimum for a tensor field, 
the solutions of the model break this symmetry.

A systematic approach to studying the theory and phenomenology 
of hypothetical diffeomorphism 
and local Lorentz violation 
is the use of effective field theory \cite{Kostelecky:1995,Colladay:1997,Colladay:1998}.
In this approach generic symmetry breaking terms are added to the action of 
General Relativity and the Standard Model \cite{Kostelecky:2004}.
These terms can generally be written using coordinate invariant contractions 
of background tensor fields and gravitational or matter field operators.
The imposition of perfect symmetry occurs when these background tensor fields vanish, 
and any statistically significant nonzero value for their components would indicate
a breaking of diffeomorphism and local Lorentz symmetry.
The use of a standard Sun-centered coordinate system to express the experimental measurements 
of these background coefficients allows cross comparisons between a wide variety of tests \cite{Kostelecky:2008}.  
For example, 
the results of time-delay VLBI measurements in the solar system already limited the coefficient $\sb_{00}$ that describes the variation of the speed of gravity 
compared to light at parts in $10^5$ \cite{LePoncinLafitte:2016}, 
well beyond initial constraints obtainable from 
the first gravitational wave observation.
Later, 
of course, 
these measurements were improvement from the simultaneous observations 
of electromagnetic observations and gravitational wave observations \cite{Abbott:2017}.
In both tests,
the same set of coefficients $\sb_\mn$ controls the deviation from GR, 
at least in the lowest order description 
of the effective field theory framework \cite{Bailey:2006,Bailey:2009,Kostelecky:2015,Kostelecky:2016}.

So far, 
much of the development of effective field theory approaches to diffeomorphism and Lorentz breaking
have focused on the weak-field regime, 
in which the spacetime metric can be expanded around a flat background
\cite{Bailey:2006,Kostelecky:2009a,Kostelecky:2011}.
For many measurements it is appropriate to use this limit 
\cite{Bourgoin:2016, Mueller:2008, Flowers:2017, Shao:2018, Hees:2015,LShao:2014}.
However, 
it is of importance to extend results beyond this level, 
and some work already exists in this regard \cite{Bailey:2016,Bonder:2015,Bonder:2017,Bailey:2019,Bonder:2020,O'Neal:2021,Xu:2020,Xu:2021,Bonder:2021}.
A generic action quadratic in the metric fluctuations $h_\mn$
has already been constructed to describe
any Lorentz violation that can be written in the form
${\cal L} \sim h {\hat K} h$, 
where ${\hat K}$ contains arbitrary coefficients and any number of partial 
derivatives \cite{Bailey:2006,Kostelecky:2016,Kostelecky:2018}.
This form assumes that the coefficients are constant and the fluctuation
around their background values can be eliminated from the action by
inserting their solutions from the field equations. 
The class of models in which this ``decoupling" is possible
has been studied in various toy models with vectors and tensors \cite{Seifert:2009,Altschul:2010,Seifert:2018}.
However, a full study remains incomplete at higher order 
in small fluctuations $h_\mn$ around a flat background.

We start the investigation in this work of higher-order terms 
in the weak-field expansion around a flat background.
A general cubic action is constructed for the case of two-tensor couplings $s_\mn$
under the assumption of spontaneous-symmetry breaking.
Using an expansion around the background or vacuum values, 
we show that when proceeding to cubic order in the action, 
terms that include the fluctuations $\st_\mn$ 
are needed for proper adherance to the underlying diffeomorphism symmetry.
Thus one cannot generally decouple the fluctuations from the metric fluctuations 
at the level of the action to cubic and higher order, 
in contrast to past assumptions about the quadratic order terms \cite{Bailey:2006}.
We show also that these terms can be matched to a general covariant
version of the effective field theory at second and higher order in $s_\mn$ 
\cite{O'Neal:2021,Kostelecky:2021}.

The paper is organized as follows.
First we consider the terms originally proposed in Ref.\ \cite{Kostelecky:2004}, 
and argue for the inclusion of dynamical terms for the fluctuations in section 2.
In section 3, 
we perform a general construction of all possible terms in the quadratic and cubic order
Lagrange densities.
Then we discuss the match to ``covariant" terms 
in section 4.
In section 5, 
we display the field equations resulting 
from the general action.
Finally we summarize in section 6, 
and comment on generalizations of this work.
We adopt natural units in this work, 
and conventions in past articles \cite{Kostelecky:2004, Bailey:2006}.

\section{Background on the action}
\label{background}

The observer covariant form of the action for the effective field theory framework in the gravity sector is
\beq
S = \frac {1}{16\pi G_N} \int d^4x \sqrt{-g} \left( (1 - u)R+ s_\ab R^\ab 
+ t_\abgd R^\abgd \right) 
+S^\prime,
\label{act}
\eeq
where $t_\abgd$, 
$s_\ab$, and $u$ 
are the coefficient fields for Lorentz violation, 
controlling the size of the mass dimension $4$ ($M^4$) Lorentz-violating terms \cite{Kostelecky:2004} (the gravitational constant $G_N$ has mass dimension $M^{-2}$).
The term $S^\prime$ contains additional dynamical contributions to the coefficients, 
should the origin of the symmetry breaking be spontaneous.

In the spontaneous symmetry breaking scenario the following expansions are used:
\bea
g_\ab &=& \et_\ab + h_\ab,
\nonumber\\
u &=& \ub + \ut,
\nonumber\\
s_\ab &=& \sb_\ab + \st_\ab,
\nonumber\\
t_\abgd &=& \tb_\abgd + \tt_\abgd.
\label{exp}
\eea
The vacuum expectation value for the coefficients are denoted with bars, for example, for $s_\ab$ it is $\langle s_\ab \rangle = \sb_\ab$, 
and we assume constancy of these coefficients (e.g., $\prt_\mu \sb_\ab =0$).
A flat background spacetime metric is assumed so that $\et_\ab$ is constant in Cartesian coordinates and we assume the partial derivatives of $\sb_\ab$ vanish to avoid global energy-momentum conservation violation \cite{Bailey:2006}.
From here on, 
we use the flat metric $\et_\ab$ and its inverse $\et^\ab$ 
to raise and lower indices on quantities unless specified.
Note that converting expressions from covariant form to 
linear, quadratic, cubic or beyond, 
must include appropriate use of \rf{exp} \cite{Bailey:2018}.  
For example, 
for the metric inverse
\bea
g^\ab &=& \et^\ab - \et^{\al\ga} \et^{\be\de} h_\gd + \et^{\al\ga} \et^{\ep\de} \et^{\be\ze} h_{\ga\de} h_{\ep\ze} + O(h^3),
\nonumber\\
&=& \et^\ab - h^\ab + h^{\al\ga} h_{\ga}^{\pt{\ga}\be}+ O(h^3),
\label{inverse-metric}
\eea
where in the second line it is understood that the indices are raised and lowered with the Minkowski metric.

Much work has been done by assuming a quadratic order and ``decoupled" version of \ref{act}.
This implies that only the vacuum values of the coefficients
$\sb_\ab$, $\tb_\abgd$, and $\ub$ contribute 
to the gravitational effects of Lorentz violation in the weak-field limit. 
The Lagrange density for the gravity sector in this limit, 
up to overall constant scalings,
can be written as
\beq
\cL= \frac {1}{4\ka} (-h_\ab G^\ab + \sb_\ab h_\gd \cG^{\al\ga\be\de}),
\label{lag2}
\eeq
with $\ka=8\pi G_N$ and 
where the Einstein curvature tensor $G^\mn$ and the curvature double dual 
$\cG^{\al\ga\be\de}$ are linear order in $h_\mn$.
Despite the lack of fluctuations $\ut, \st, \tt$ in the second term, 
this remains invariant under linearized gauge transformations of $h_\mn$.
Thus the fluctuations of the coefficients $\ut$, $\st_\ab$, 
and $\tt_\abgd$ are assumed to have been decoupled via their field equations.
The latter process involves assumptions on the dynamics in $S'$ 
\cite{Bailey:2006}, 
which limits the class of possible models included \cite{Seifert:2018}.
On the other hand, 
\rf{lag2} represents the simplest limit 
in which the focus is on the effects acting on 
only the known gravitational degrees of freedom in $h_\mn$, 
thereby invoking Occam's razor for the 
already ambitious suggestion of Lorentz violation.

Note also that the coefficients $\tb_\abgd$ do not appear in the quadratic action \rf{lag2}.
That this occurs can be viewed as a consequence of the assumptions on the dynamics
of the coefficients that leads to the decoupled form of \rf{lag2}.
Alternatively, 
\rf{lag2} can be derived by considering all possible operators
in a general quadratic action of the form $h K \prt \prt ... h$ \cite{Kostelecky:2016}, 
and then truncating the expansion to only mass dimension $4$ coefficients for Lorentz violation.
Either way, 
a deeper understanding of why the $\tb_\abgd$ coefficients vanish in this limit
represents a statement of the ``t-puzzle" \cite{Bailey:2015}.
While alternative covariant actions have been countenanced leading 
to physical effects of $t_\abgd$ \cite{Bonder:2017,Bonder:2020,Bonder:2021}, 
it remains unexplained why it does not occur in \rf{lag2}.

In this work, 
we include the fluctuation terms so that we relax
the assumptions made in \rf{lag2},
and thus consider a more general version of $S^\prime$.
To include the generality of possible $\st_\mn$ dynamics, 
we posit all possible kinetic, potential, 
and coupling terms with the constraint that we consider only terms 
in the Lagrange density with no higher than second order in derivatives, 
and first-order terms in the vacuum coefficients $\sb_\mn$.
We will attempt here the $s_\mn$ case but other coefficients can similarly be accommodated.

\section{General construction of Lagrange density}
\label{general}

First we start with the quadratic action terms.
These Lagrange density terms, 
discarding those equivalent by partial integration, 
are constructed from scalar density contractions of the following:
\bea
\cL^{(2)}  &\supset& {\rm All \, contractions \, of\,} \{ 
h_\ab \prt_\ga \prt_\de h_{\ep\ze},
\sb_\ab h_\gd \prt_\ep \prt_\ze h_{\et\th}, 
\st_\ab \prt_\ga \prt_\de \st_{\ep\ze}, 
\st_\ab \prt_\ga \prt_\de h_{\ep\ze}, 
\nonumber\\
&& \pt{all contractions of}
\sb_\ab \st_\gd \prt_\ep \prt_\ze h_{\et\th},
\sb_\ab \st_\gd \prt_\ep \prt_\ze \st_{\et\th}
 \}.
\label{QlagForm}
\eea
In addition there are potential-type terms dealt with separately below.
Note that the form of the terms considered in \rf{QlagForm} includes, 
in the second set of terms $\sim \sb h \prt \prt h$, 
the general structure of the Lorentz-violating terms considered 
in the quadratic lagrangian construction of Ref.\ \cite{Kostelecky:2016}, 
and it will include the General Relativity terms from $\sim h \prt \prt h$.
We truncate the series in this work by considering terms of second order in derivatives, 
though the pattern could be generalized to accommodate higher derivative terms.

For the cubic terms in the action, 
the construction follows similarly by including all possible terms cubic in $h_\ab$, 
or quadratic in $h_\ab$ and linear in $\st_\ab$, etc.
We truncate the expansion to leading order in the vacuum values $\sb_{ab}$, 
since these terms will be of primary interest for phenomenology.
The structure of these terms, 
again discarding those equivalent by integration by parts,
is
\bea
\cL^{(3)}  &\supset& {\rm All \, Contractions \, of\,} \{ 
h_\ab h_{\ga\de} \prt_\ep \prt_\ze h_{\et\th},
h_\ab \prt_\ga h_{\de\ep} \prt_\ze h_{\et\th},
\sb_\ab h_\gd h_{\ep\ze} \prt_\et \prt_\th h_{\ka\la}, 
\sb_\ab h_\gd \prt_\ep h_{\ze \et} \prt_\th h_{\ka\la}, 
\nonumber\\
&&
h_\ab \st_\gd \prt_\ep \prt_\ze \st_{\th\ka},
h_\ab \prt_\ga \st_{\de\ep} \prt_\ze \st_{\th\ka},
\sb_\ab \st_\gd h_{\ep\ze} \prt_\et \prt_\th h_{\ka\la},
\sb_\ab \st_\gd \prt_\ep h_{\ze \et} \prt_\th h_{\ka\la},
\st_\ab h_\gd \prt_\ep \prt_\ze h_{\et\th},
\nonumber\\
&&
\st_\ab \prt_\ga h_{\de\ep} \prt_\ze h_{\et\th},
\st_\ab \st_\gd \prt_\ep \prt_\ze \st_{\et\th},
\st_\ab \prt_\ga \st_{\de\ep} \prt_\ze \st_{\et\th},
\sb_\ab h_\gd \st_{\ep\ze} \prt_\et \prt_\th \st_{\ka\la},
\sb_\ab h_\gd \prt_\ep \st_{\ze\et} \prt_\th \st_{\ka\la}
\}.
\label{ClagForm}
\eea
Among all the terms considered in \rf{QlagForm} and \rf{ClagForm}
are the would-be decoupled terms, i.e., those terms that do not have $\st$ in them.
If such terms are possible through cubic order, they will appear in this construction.

We next impose the requirement of diffeomorphism invariance of the action 
without assuming the field equations are satisfied (this is an ``off-shell" requirement).
The quantity constructed by the series
\beq
\cL= \cL^{(2)} + \cL^{(3)} + ...,
\label{series}
\eeq 
is a scalar density of weight $-1$.  
In the spontaneous-symmetry breaking scenario,
this function depends on the matter fields,
the vacuum values of the coefficients for Lorentz violation ($\sb_\mn$), 
the metric fluctuations $h_\mn$ and coefficient fluctuations $\st_\mn$.
To impose the symmetry requirement, 
we will need the infinitesimal diffeomorphism symmetry transformations, 
which are given by the Lie derivatives of the underlying fields $g_\mn$ and $s_\mn$ 
along an arbitrary vector field $\xi^\mu$.  
These transformations are
\bea
\cL_\xi g_\mn &=& g_{\la\mu} \prt_\nu \xi^\la  + g_{\la\nu} \prt_\mu \xi^\la + \xi^\la \prt_\la g_\mn,
\nonumber\\
\cL_\xi s_\mn &=& s_{\la\mu} \prt_\nu \xi^\la  + s_{\la\nu} \prt_\mu \xi^\la + \xi^\la \prt_\la s_\mn.
\label{Lie}
\eea
From this we can derive the effective Lie derivatives, 
or diffeomorphism transformations for the field fluctuations:
\bea
\cL_\xi h_\mn &=& \prt_\mu \xi_\nu + \prt_\nu \xi_\mu + h_{\mu\la} \prt_\nu \xi^\la  + h_{\nu\la} \prt_\mu \xi^\la + \xi^\la \prt_\la h_\mn,
\nonumber\\
\cL_\xi \st_\mn &=&  (\sb_{\la\mu}+\st_{\la\mu}) \prt_\nu \xi^\la  + (\sb_{\la\nu}+\st_{\la\nu}) \prt_\mu \xi^\la + \xi^\la \prt_\la \st_\mn.
\label{Lie2}
\eea
where we impose $\prt_\la \sb_\mn=0$ in the chosen asymptotically flat cartesian coordinates.
In this treatment, 
it is necessary to keep all terms to linear order in $\xi_\al$, 
in order to result in the correct conservation laws.
This includes terms typically discarded like $\sim \xi h$ in the linearized limit.

The total action will take the form
\beq
S=S_M+ \frac {1}{2 \ka} \int d^4x ( \cL^{(2)} + \cL^{(3)} + ...),
\label{act2}
\eeq
where the first term is the matter sector, 
which we assume to be minimally coupled to gravity.
The remaining terms are those that we seek to construct 
and are the focus in what follows.
Under a diffeomorphism, 
one can show the total change in the action takes the form 
\beq
\de S = \frac {1}{2\ka} \int d^4 x \left( \fr {\de \cL}{\de h_\ab} \cL_\xi h_\ab + \fr {\de \cL}{\de \st_\ab} \cL_\xi \st_\ab \right).
\label{deltaS}
\eeq
At this stage we make no assumption about whether the field equations are satisfied.
Insertion of \rf{Lie2} and using integration by parts gives
\bea
\de S &=& -\frac {1}{2\ka} \int d^4 x \bigg\{
\prt_\be \left( \fr {\de \cL}{\de h_{\ga\be}} \right) + \Ga^\ga_{\pt{\be}\al\be} \fr {\de \cL}{\de h_\ab} 
+  g^{\de\ga} s_{\de\al} \prt_\be \left( \fr {\de \cL}{\de \st_\ab} \right)
\nonumber\\
&&
\pt{\frac {1}{4\ka} \int d^4x }
 +  g^{\de\ga} \frac 12 (\prt_\al \st_{\be\de}+\prt_\be \st_{\al\de} -\prt_\de \st_\ab) \fr {\de \cL}{\de h_{\al\ga}} \bigg\} g_{\ga\ep} \xi^\ep.
\label{deltaS2}
\eea
The quantity in braces will vanish for ``off-shell" diffeomorphism invariance 
of the action in the spontaneous-symmetry breaking scenario.
For more details, 
the reader is referred to expositions on Noether identities 
in the context of Lorentz and diffeomorphism breaking presented 
in Refs.\ \cite{Kostelecky:2004,Bluhm:2005,Bluhm:2008,Bluhm:2015,Bluhm:2016, Bluhm:2019, Bluhm:2021}.

We take the general expansion 
in equations \rf{QlagForm} and \rf{ClagForm} and insert an arbitrary parameter 
in front of each independent contraction.
The starting lagrange density for quadratic order takes the form
\beq
\cL^{(2)} = C_1 h^\ab \sb^\gd \prt_\al \prt_\be h_\gd + C_2 h^\ab \sb^\ga_{\pt{\ga}\ga} \prt_\al \prt_\be h^\de_{\pt{\de}\de} 
+ C_3 h^\ab \sb_\al^{\pt{\al}\ga} \prt_\ga \prt_\be h^\de_{\pt{\de}\de}+...,
\label{QlagEx}
\eeq
with the remaining terms not displayed here for brevity.
For the cubic terms, these take the form
\beq
\cL^{(3)} = A_1 h_\al^{\pt{\al}\ga} h^\ab \sb^{\de\ep} \prt_\ga \prt_\be h_{\de\ep} + A_2 h h^\bg \sb^{\de\ep} \prt_\ga \prt_\be h_{\de\ep} 
+ A_3 h_\al^{\pt{\al}\ga} h^\ab \sb^\de_{\pt{\de}\de} \prt_\ga \prt_\be h + ...,
\label{ClagEx}
\eeq
displaying just the first few terms.
Note that because the Lagrange density has mass dimension $4$, 
and $s_\mn$ is taken as dimensionless,
the parameters $C_n$ and $A_n$ have mass dimension $2$.

The idea is to insert the total lagrange density into the vector combination of terms in 
brackets \rf{deltaS2} above, 
and demand the result be zero.
All tensorially independent terms must vanish and this imposes many relations among the arbitrary parameters $\{C_n \}$ and $\{A_n \}$.
When one imposes these relations on the parametrized lagrange densities \rf{QlagEx} and \rf{ClagEx}, 
a vastly smaller subset of these terms remains, 
each obeying the off-shell conservation law expected from the underlying diffeomorphism symmetry.
This entire calculation is carried out within the Mathematica set of packages in {\it xAct}, 
including in particular, the {\it xTras} package \cite{xTras:2014}.

We now display the Lagrange density result in several pieces.
Firstly, 
the entire process produces the quadratic and cubic lagrange densities of General Relativity, 
up to integration by parts.
Secondly, 
scalings of GR by the trace $\sb^\al_{\pt{\al}\al}$ occur,
which we leave out of our tables below because these are unobservable scalings, 
as expected. 
Neglected also are terms that satisfy the constraints \rf{deltaS2}, 
but do not contribute to the gravitational field equation for $h_\mn$
like, for example, terms in equation \rf{QlagForm} of the form $\sb \prt \st \prt \st$.
We find the following quadratic terms organized in the generic form
\beq
\cL^{(2)} = \cL^{(2)}_{GR} + \sum^{10}_{n=1} \al_n \cM^{(2)}_n,
\label{Lform}
\eeq
where the $\cM^{(2)}_n$ terms are listed in the Table \ref{M2}.
Where necessary for clarity, 
terms with the parentheses label of $(1)$ are first order in the metric $h_\ab$ and $\st_\ab$ fluctuations while those with the label $(2)$ are second order.
All Christoffel symbols are taken to first order in $h_\ab$ 
with indices raised and lowered with the flat metric.
It is also useful to use a linearized Christoffel symbol $\Gat$ with the substitution $h_\ab \rightarrow \st_\ab$.
Thus
\bea
\Ga_{\al\be\ga} &=& \frac 12 (\prt_\be h_{\ga\al}+ \prt_\ga h_{\be\al} 
- \prt_\al h_{\be\ga}  ),\nonumber\\
\Gat_{\al\be\ga} &=& \frac 12 (\prt_\be \st_{\ga\al}+ \prt_\ga \st_{\be\al} 
- \prt_\al \st_{\be\ga}  ). 
\label{gat}
\eea
Two different contractions of the Christoffel symbols with one free index are defined by
\bea
\Ga_{A\mu} &=& \Ga^\al_{\pt{\al}\al\mu},
\nonumber\\
\Ga_B^\mu &=& \Ga^{\mu\al}_{\pt{\mu\al}\al},
\label{gammas}
\eea
with the same definitions holding for $\Gat$.
We employ some shorthands like $\Ga_A^2=(\Ga_A)^\al (\Ga_A)_\al$, $h=h_\ab \et^\ab$, and $\st=\st_\ab \et^\ab$.
Finally, 
we use shorthands for curvature-like quantities built from $\Gat_{\al\be\ga}$, 
like, 
for example, 
$\Gt^{(1)}_\mn = \prt_\al \Gat^\al_{\pt{\al}\mu\nu} - \prt_\mu \Gat_{A\nu} -\fr 12 \et_\mn 
( \prt^\al \Gat_{B\al} - \prt^\al \Gat_{A\al} )$.

\begin{table}[h]
\centering
\begin{tabular}{| l | c |}
\hline
Term & Expression  \\ [0.5ex]
\hline
$\cM^{(2)}_{\rm GR}$ & $-\tfrac {1}{4\ka} h^\ab G^{(1)}_\ab$ \\ 
\hline
$\cM^{(2)}_1$ & $\sb_\ab \st^\ab R^{(1)} $ \\ 
\hline
$\cM^{(2)}_2$ & $\st^\ab \Gt^{(1)}_\ab - \frac 12 \st \prt^\be \Gat_{A\be} + \st \, \sb^\ab \prt_\al \Ga_{B \be} $ \\ 
\hline
$\cM^{(2)}_3$ & $\sb_\al^{\pt{\al}\be} \st^{\al\ga}( R^{(1)}_{\be\ga} - \prt^\de \Ga_{\be\ga\de} ) + \frac 12 \st^\ab \prt^\ga \Gat_\abg $ \\ 
\hline
$\cM^{(2)}_4$ & $-\st^\ab \Gt^{(1)}_\ab$ \\ 
\hline
$\cM^{(2)}_5$ & $ \sb^\ab \st_\al^{\pt{\al}\ga} ( R^{(1)}_{\be\ga} + \prt^\de \Ga_{\be\ga\de} ) - \frac 12 \st^\ab \prt^\ga \Gat_{\al\be\ga}$ \\  
\hline
$\cM^{(2)}_6$ & $ 2 \sb^\ab \st^\gd (R^{(1)}_{\ga\al\be\de} 
- \prt_\de \Ga_{\al\be\ga} ) + \st^\ab \prt_\ga \Gat^\ga_{\pt{\ga}\al\be}$ \\ 
\hline
$\cM^{(2)}_7$ & $ -\st^\ab \prt_\ga \Gat^\ga_{\pt{\ga}\al\be} + 2 \sb^\ab \st^\gd \prt_\de \Ga_\abg$ \\ 
\hline
$\cM^{(2)}_8$ & $-\st \prt^\al \Gat_{A\al} + 2 \sb^\ab \st \prt^\de \Ga_{\al\be\de}$ \\ 
\hline
$\cM^{(2)}_9$ & 
$ h^\ab \sb^\gd \cG^{(1)}_{\al\ga\be\de}  - \sb^\ab h_\ab R^{(1)} + 2 \st^\ab R^{(1)}_\ab $\\ 
\hline
$\cM^{(2)}_{10}$ & 
$ (\st - \sb^\ab h_\ab) R^{(1)}$\\ 
\hline
\end{tabular}
\caption{\label{M2}Quadratic order terms in \rf{Lform}.}
\end{table}

Similarly, 
the cubic terms can be organized as
\beq
\cL^{(3)} = \sum^{10}_{n=1} \al_n \cM^{(3)}_n,
\label{Lform3}
\eeq
and they are listed in Table \ref{M3}.

\begin{table}[h]
\centering
\begin{tabular}{| l | c |}
\hline
Term & Expression  \\ [0.5ex]
\hline
$2 \ka \cM^{(3)}_{\rm GR}$ 
& $h^\ab (\Ga_{\al\be}^{\pt{\al\be}\ga} \Ga_{B\ga} +\Ga^\ga_{\pt{\ga}\al\be}\Ga_{A\ga} 
-\Ga^{\ga\de}_{\pt{\ga\de}\al} \Ga_{\de\ga\be} -3 \Ga_{\al\ga\de} \Ga^{\ga\de}_{\pt{\ga\de}\be}
+\Ga_{A\al} \Ga_{B\be} + \fr 12 \Ga_{\al\be}^{\pt{\al\be}\ga} \Ga_{A\ga} 
- \fr 12 \Ga^{\ga\de}_{\pt{\ga\de}\al} \Ga_{\ga\de\be} - \fr 12 \Ga_{\al\ga\de} \Ga_\be^{\pt{\be}\ga\de} )$ \\
& $+h (\fr 34 \Ga^{\al\be\ga} \Ga_{\ga\al\be} 
- \fr 12 \Ga_{A\al} \Ga_B^\al +\fr 14 \Ga^{\al\be\ga} \Ga_{\al\be\ga}) 
+\fr 12 (\fr 12 h h^\ab - h^{\al\ga} h^\be_{\pt{\al}\ga} ) \prt^\de \Ga_{\al\be\de} $ \\ 
\hline
$\cM^{(3)}_1$ 
& $\fr 12 \st_\ab \st^\ab R^{(1)} - 2 h^\ab \sb_\al^{\pt{\al}\ga} \st_{\be\ga} R^{(1)} 
+ \sb_\ab \st^\ab (\Ga_{\al\be\ga} \Ga^{\al\be\ga} 
- \Ga_B^2 + \fr 12 h R^{(1)} -2 h^\ab R^{(1)}_\ab )$  \\ 
\hline
$\cM^{(3)}_2$ 
& $2 \sb^\ab \st^\gd ( \Ga_{\al\ga\de} \Ga_{B \be} - \Ga_{\al\ga}^{\pt{\al\ga}\et} \Ga_{\be\de\et} ) 
- \sb^\ab \st (2 \Ga_\al^{\pt{\al}\de\et} \Ga_{\de\et\be} 
+\Ga_{B\de} \Ga_\ab^{\pt{\al\be}\de} 
+ \Ga_{B\de} \Ga^\de_{\pt{\de}\al\be} +  \Ga_{B\al} \Ga_{B\be})$ \\
& $+ h^\ab (2 \Gat_\al^{\pt{\al}\ga\de} \Gat_{\ga\de\be} 
-\Gat_\al^{\pt{\al}\ga\de} \Gat_{\be\ga\de}
+ \Gat^{\ga\de}_{\pt{\ga\de}\al} \Gat_{\de\ga\be}
- 2 \Gat_{\al\be}^{\pt{\al\be}\ga} \Gat_{B \ga} 
+ \Gat_{B\al} \Gat_{B\be} ) 
- h ( \Gat^{\al\be\ga} \Gat_{\be\al\ga} + \frac 12 \Gat_A^2  ) $\\
& $ + (\frac 12 h \st \sb^\ab - 2 \st h^{\ga\al} \sb_\ga^{\pt{\ga}\be}   
- h^\gd \st_\gd \sb^\ab ) ( R^{(1)}_\ab + \prt^\et \Ga_{(\al\be)\et} )
+2 h^\ab \st_\al^{\pt{\al}\ga} \prt_\ga \Gat_{A \be} - h \st^\ab \prt_\al \Gat_{A\be} 
-h^\ab \sb^\gd \st \prt_\de \Ga_{\ga\al\be}$ \\
\hline
$\cM^{(3)}_3$ 
& $\sb^\ab \st^\gd \Ga_{\al\ga\ep} \Ga_{\be\de}^{\pt{\be\de}\ep} 
+ \sb^\ab \st_\al^{\pt{\al}\ga} 
(2 \Ga_{\be\de\ep} \Ga^{\de\ep}_{\pt{\de\ep}\ga} + 2 \Ga_{\de\ep\ga} \Ga^{\de\ep}_{\pt{\de\ep}\be} 
+ \Ga_{\be\ga\de} \Ga_B^\de - \Ga_{B\de} \Ga^\de_{\pt{\de}\be\ga}) $\\
& $ +h^\ab ( \frac 12 \Gat^\gd_{\pt{\ga\de}\al} \Gat_{\de\ga\be} - \frac 12 \Gat_{\al\ga\de} \Gat_\be^{\pt{\be}\ga\de} - \Gat_{\al\ga\de} \Gat^{\ga\de}_{\pt{\ga\de}\be}
+\Gat_{\al\be\ga} \Gat^\ga_B + \Gat_{B \ga} \Gat^\ga_{\pt{\ga}\al\be} + \Gat_\ab^{\pt{\al\be}\ga} \Gat_{A\ga} + \Gat^\ga_{\pt{\ga}\al\be} \Gat_{A\ga} )$\\
& $ - h (\frac 12 \Gat_{\be\ga\de} \Gat^{\be\ga\de} + \Gat_{\be\ga\de} \Gat^{\ga\de\be} 
+ \frac 14 \Gat_B^2  + \frac 12 \Gat_{A\be} \Gat_B^\be + \frac 14 \Gat_A^2 )
+ h^\ab \sb^\gd \st_\ga^{\pt{\ga}\ep} (\prt_\al \Ga_{\de\be\ep} + R^{(1)}_{\de\al\be\ep})$\\
& $ +h^\ab \sb^\gd \st_{\al\ga} (\prt^\ep \Ga_{\de\be\ep} - R^{(1)}_{\be\de} ) 
+\frac 12 h \sb^{\be\ga} \st_\be^{\pt{\be}\de} (R^{(1)}_\gd - \prt^\ep \Ga_{\ga\de\ep} )
- \frac 12 h \st^{\be\ga} (\prt^\de \Gat_{\be\ga\de} + \prt_\de \Gat^\de_{\pt{\de}\be\ga})$ \\
& $ +h^\ab \sb_{\al\ga} \st^\gd ( \prt^\ep \Ga_{\be\de\ep}-R^{(1)}_{\be\de})
+h^\ab \sb_\al^{\pt{\al}\ga} \st_\be^{\pt{\be}\de} ( \prt^\ep \Ga_{\ga\de\ep}-R^{(1)}_{\ga\de} )
+ 2 h^\ab ( \st^\gd \prt_\de \Gat_{(\al\ga)\be} - \st_\al^{\pt{\al}\ga} \prt^\de \Gat_{(\be\ga) \de} )$\\
\hline
$\cM^{(3)}_4$ 
& $-h^\ab \sb_\ab \st^\gd R^{(1)}_\gd 
+ 2 \sb^\ab \st^\gd ( \Ga_{\al\ga\ep} \Ga_{\be\de}^{\pt{\be\ga}\ep}  - \Ga_{B\al} \Ga_{\be\ga\de})
+\sb^\ab \st (\Ga_{B\al} \Ga_{B\be} - \Ga_{\al\de\ep} \Ga_\be^{\pt{\be}\de\ep})
+h ( \Gat_\bgd \Gat^{\ga\de\be} - \Gat_{A\be} \Gat_B^\be )
$\\
& $ +\st \Ga_{A\be} (\Gat_A^\be + \Gat_B^\be) 
+ h^\ab ( \Gat_{A\al} \Gat_{A\be} + 2 \Gat_{B\al} \Gat_{A \be} + 2 \Gat^\ga_{\pt{\ga}\ab} \Gat_{A \ga} - 2 \Gat_\ab^{\pt{\al\be}\ga} \Gat_{A\ga} - \Gat_{B\al} \Gat_{B\be} $\\
&$+ 2 \Gat_\ab^{\pt{\al\be}\ga} \Gat_{B\ga} - 2 \Gat_{\ga\de\al} \Gat_\be^{\pt{\be}\ga\de} - \Gat_{\ga\de\al} \Gat^{\de\ga}_{\pt{\de\ga}\be} + \Gat_{\al\ga\de} \Gat_\be^{\pt{\be}\ga\de} 
+\st \prt^\de \Gat_{\de\al\be})$\\ 
\hline
$\cM^{(3)}_5$ 
& $-\sb^\ab \st^\gd \Ga_{\al\ga}^{\pt{\al\ga}\ep} \Ga_{\be\de\ep} - \sb^\ab \st_\al^{\pt{\al}\ga} ( 2 \Ga_\be^{\pt{\be}\de\ep} \Ga_{\de\ep\ga}  + \Ga_{\be\ga\de} \Ga_B^\de + \Ga_B^\de \Ga_{\de\be\ga})
+ h^\ab ( \Gat_\al^{\pt{\al}\ga\de} \Gat_{\ga\de\be} - \frac 12 \Gat_{\al\ga\de} \Gat_\be^{\pt{\be}\ga\de} + \frac 12 \Gat^{\ga\de}_{\pt{\ga\de}\al} \Gat_{\de\ga\be} )$ \\ 
& $ -\frac 12 h ( \Gat^{\be\ga\de} \Gat_{\ga\de\be} + \frac 12 \Gat_B^2 + \Gat_{A\ga} \Gat^\ga_B + \frac 12 \Gat_{A}^2 )
- h^\ab \sb^\gd \st_{\al\ga} (R^{(1)}_{\be\de} + \prt^\ep \Ga_{\de\be\ep}) 
- h^\ab \sb_\al^{\pt{\al}\ga} \st_\ga^{\pt{\ga}\de} ( \prt^\ep \Ga_{\be\de\ep} 
+ R^{(1)}_{\be\de} )  $\\ 
& $ + \frac 12 h \sb^{\be\ga} \st_\be^{\pt{\be}\de} 
( \prt^\ep \Ga_{\ga\de\ep} - \prt_\de \Ga_{A\ga}) 
- \frac 12 h \st^{\be\ga} ( \prt^\de \Gat_{\be\ga\de} + \prt_\de \Gat^\de_{\pt{\de}\be\ga} ) 
+ h^\ab \st_\al^{\pt{\al}\ga} (\prt^\de \Gat_{\ga\be\de} + \prt_\de \Gat^\de_{\pt{\de}\be\ga} )$\\
& $ - h^\ab \sb_\al^{\pt{\al}\ga} \st_\be^{\pt{\be}\de} 
(R^{(1)}_{\ga\de} + \prt^\ep \Ga_{\ga\de\ep} )
-h^\ab \sb^{\ga\de} \st_\ga^{\pt{\ga}\ep} \prt_\ep \Ga_{\de\al\be} + \frac 12 h \sb^{\be\ga} \st_\be^{\pt{\be}\de} \prt_\ep \Ga^\ep_{\pt{\ep}\ga\de}
$\\
\hline
$\cM^{(3)}_6$ 
& $2 \sb^\ab \st^\gd ( 2 \Ga_{\al\ga\ep} \Ga^\ep_{\pt{\ep}\be\de} + \Ga_{\al\be}^{\pt{\al\be}\ep} \Ga_{\ep\ga\de} + \Ga_{\ep\ga\de} \Ga^\ep_{\pt{\ep}\al\be} + \Ga_{\al\ga\de} \Ga_{B\be} )
+2 h^\ab ( \Gat_{\al\be}^{\pt{\al\be}\ga} \Gat_{B\ga} + \Gat_{\al\be}^{\pt{\al\be}\ga} \Gat_{B\ga} - \Gat_\al^{\pt{\al}\ga\de} \Gat_{\be\ga\de})$ \\ 
& $ -h \Gat^{\be\ga\de} \Gat_{\ga\de\be} +2 h^\ab \sb_\al^{\pt{\al}\ga} \st^{\de\ep} (2 R^{(1)}_{\de\be\ep\ga} + \prt_\ep \Ga_{\be\ga\de}+ \prt_\ep \Ga_{\ga\de\be})
+h \sb^{\be\ga} \st^{\de\ep} (R^{(1)}_{\de\be\ga\ep}  - \prt_\ep \Ga_{\be\ga\de})$\\
& $+2 h^\ab \sb^\gd \st_\al^{\pt{\al}\ep} ( R^{(1)}_{\be\ga\ep\de} + \prt_\de \Ga_{\ga\be\ep}+ \prt_\ep \Ga_{\ga\be\de} )
+2 h^\ab \st^\gd \prt_\de \Gat_{\al\be\ga}  - 2 h^\ab \st_\al^{\pt{\al}\ga} (\prt_\de \Gat_{\be\ga}^{\pt{\be\ga}\de} + \prt_\de \Gat^\de_{\pt{\de}\be\ga} )$\\
\hline
$\cM^{(3)}_7$ 
& $ -2 \sb^\ab \st^\gd (2 \Ga_{\al\ga}^{\pt{\al\ga}\ep} \Ga_{\ep\be\de} + \Ga_{\al\be}^{\pt{\al\be}\ep} \Ga_{\ep\ga\de} 
+ \Ga_{\ep\be\de} \Ga^\ep_{\pt{\ep}\al\ga} + \Ga_{\al\ga\de} \Ga_{B\be} )$\\
& $+h^\ab ( \Gat_\al^{\pt{\al}\ga\de} \Gat_{\be\ga\de} 
-2 \Gat_\al^{\pt{\al}\ga\de} \Gat_{\ga\de\be} - \Gat^{\ga\de}_{\pt{\ga\de}\al} \Gat_{\de\ga\be}
+2 \Gat_{\al\be}^{\pt{\al\be}\ga} \Gat_{B\ga} 
+ 2 \Gat_{\al\be}^{\pt{\al\be}\ga} \Gat_{A\ga} - \Gat_{B\al} \Gat_{B\be} -2 \Gat_{B\al} \Gat_{A\be} - \Gat_{A\al} \Gat_{A\be}  )$\\
& $ + h \Gat_{\al\be\ga} \Gat^{\be\ga\al} + h^\ab \st_\ab ( \prt_\ga \Gat^\ga_B + \prt_\ga \Gat^\ga_A  ) -4 h^\ab \sb_\al^{\pt{\al}\ga} \st^{\de\ep} \prt_\ep \Ga_{(\be\ga)\de} 
+ h \sb^\ab \st^\gd \prt_\de \Gat_{\al\be\ga} - 4 h^\ab \sb^\gd \st_\al^{\pt{\al}\ep} \prt_\ep \Ga_{\ga\be\de}$\\
\hline
$\cM^{(3)}_8$ 
& $ 4 \sb^\ab \st^\gd \Ga_{\al\be}^{\pt{\al\be}\ep} \Ga_{\ga\de\ep} + 4 h^\ab \st^\gd (\Ga_{\ga\de}^{\pt{\ga\de}\ep} \Gat_{\al\be\ep} + \Ga_{\ga\al\de} \Gat_{A\be} )
- 2 h^\ab ( 2 \Gat_{\al\be}^{\pt{\al\be}\ga} \Gat_{A\ga} + \Gat_{A\al} \Gat_{A\be} )$\\
& $ -4 \st^\ab \Ga_\ab^{\pt\ab \ga} \Gat_{A\ga} + h \Gat_A^2 - 2 h \sb^\ab \Ga_\ab^{\pt\ab \ga} \Gat_{A\ga} 
+ 8 h^\ab \sb_\al^{\pt{\al}\ga} \Ga_{(\be\ga)}^{\pt{(\be\ga)} \de} \Gat_{A\de} $\\
\hline
$\cM^{(3)}_9$ 
& $ h^\ab \sb^\gd ( 4 \Ga_{\al\ga}^{\pt{\al\ga}\ep} \Ga_{[ \be\de ] \ep} + 2 \Ga_\ab^{\pt\ab \ep} \Ga_{\ga\de\ep} - 3 \Ga_{\ga\al}^{\pt{\ga\al}\ep} \Ga_{\de\be\ep}
-2 \Ga_{\ga\de}^{\pt\gd \ep} \Ga_{\ep\al\be} + 2 \Ga_{\al\ga}^{\pt{\al\ga}\ep} \Ga_{\ep\be\de} + 2 \Ga_\ab^{\pt\ab \ep} \Ga_{\ep\ga\de}$\\
& $ -2 \Ga_{\ep\ga\de} \Ga^\ep_{\pt{\ep}\al\be} + \Ga_{\ep\be\de} \Ga^\ep_{\pt{\ep}\al\ga} - 8 \Ga_{(\al\ga)\de} \Ga_{A\be} + 4 \Ga_{\ga\al\be} \Ga_{A\de} 
- 8 \Ga_{(\al\ga)\de} \Ga_{B\be} )$\\
& $ +h^\ab \sb_\al^{\pt{\al}\ga} ( 2 \Ga_{\de\ep\ga} \Ga^{\de\ep}_{\pt{\de\ep}\be} + 2 \Ga^{\de\ep}_{\pt{\de\ep}\be} \Ga_{\ep\de\ga} - 2 \Ga_{B\be} \Ga_{B\ga}
-6 \Ga_{(\be\ga)}^{\pt{(\be\ga)}\de} \Ga_{B\de} -4 \Ga_{B\de} \Ga^\de_{\pt{\de}\be\ga} - 2 \Ga_{B\ga} \Ga_{A\be} $\\
& $ + 4 \Ga_{B\be} \Ga_{A\ga} - 8 \Ga_{(\be\ga)}^{\pt{(\be\ga))}\de} \Ga_{A\de} - 8 \Ga^\de_{\pt{\de}\be\ga} \Ga_{A\de})$\\
& $ + h \sb^\ab (-\Ga_{(\de\ep)\be} \Ga^{\de\ep}_{\pt{\de\ep}\al} + \frac 12 \Ga_{B\al} \Ga_{B\be} + \Ga_\ab^{\pt\ab \ga} \Ga_{B\ga} + \Ga_{B\ga} \Ga^\ga_{\pt{\ga}\al\be}
+\Ga_{B\al} \Ga_{A\be} + \frac 32 \Ga_{A\al} \Ga_{A\be} -\Ga_\ab^{\pt\ab \ga} \Ga_{A\ga} - \Ga^\de_{\pt{\de}\al\be} \Ga_{A\de})$\\
& $ + 2 \st^\ab ( \Ga_{\ga\de\be} \Ga^{\ga\de}_{\pt{\ga\de}\al} - \Ga_{B\ga} \Ga^\ga_{\pt{\ga}\al\be} )+ 2 h^\ab \st^\gd R^{(1)}_{\ga\al\be\de} - 4 h^\ab \st_\al^{\pt{\al}\ga} R^{(1)}_\bg
+ h \st^\ab R^{(1)}_\ab -2 h^\ab h^\gd \sb_\al^{\pt{\al}\ep} \prt_\de \Ga_{(\be\ep)\ga} $\\
& $- 3 h^\ab h^\gd \sb_{\al\ga} \prt_\de \Ga_{A\be} -4 h_\al^{\pt{\al}\ga} h^\ab \sb_\be^{\pt{\be}\de} \prt^\ep \Ga_{(\ga\de)\ep} + h h^\ab \sb_\be^{\pt{\be}\de} \prt^\ep \Ga_{(\ga\de)\ep}$\\
\hline
$\cM^{(3)}_{10}$ 
& $\st ( \Ga_\abg \Ga^\abg - \Ga_{B\al} \Ga_B^\al) + h^\ab \sb^\gd ( 2 \Ga_\ab^{\pt\ab \ep} \Ga_{\ga\de\ep} - \Ga_{\ga\de}^{\pt\gd \ep} \Ga_{\ep\al\be} + \Ga_{\ga\de\al} \Ga_{A\be}
- 3 \Ga_{B\al} \Ga_{\ga\de\be} ) $\\
& $-\frac 12 h^\ab \sb_\ab (\Ga_{B\ga} \Ga_B^\ga + \Ga_{A\ga} \Ga_A^\ga) 
+ 4 h^\ab \sb_\al^{\pt{\al}\ga} ( \Ga_{(\be\ga)}^{\pt{(\be\ga)}\de} \Ga_{A\de} - \Ga_{(\be\ga)}^{\pt{(\be\ga)}\de} \Ga_{B\de}  )  
+ h \sb^\ab (\Ga_\ab^{\pt\ab \ga} \Ga_{B\ga} - \Ga_\ab^{\pt\ab \ga} \Ga_{A\ga} )$\\
& $ -2 h^\ab \st R^{(1)}_\ab + (\frac 12 h \st  - h^\ab \st_\ab ) R^{(1)} - \frac 12 h^\ab h^\gd \sb_\ab \prt^\ep \Ga_{\ga\de\ep}$\\  
\hline
\end{tabular}
\caption{\label{M3}Cubic order terms in \rf{Lform3}.  Each term, $\cM^{(2)}_n+ \cM^{(3)}_n$, 
is invariant under the diffeomorphism transformations of the metric fluctuations $h_\ab$ and the fluctuations $\st_\ab$ in equations \rf{Lie} and \rf{Lie2}.}
\end{table}

One consequence of this construction is that there are no terms $\cM^{(2)}_n+ \cM^{(3)}_n$ that are devoid of the fluctuations $\st_\ab$, 
save for the GR terms.
In other words, 
in equations \rf{QlagForm} and \rf{ClagForm}, 
terms of the form $\sb h \prt \prt h$ with cubic counterparts 
$h \sb \prt h \prt h$, $h \sb h \prt \prt h$ would be decoupled, 
but such terms do not occur in the results in the Tables 1 and 2.
Furthermore, 
no linear combination of these terms can be taken to eliminate the fluctuations.
This shows it is not possible to construct terms in the action at cubic order, 
and second order in derivatives, 
that arise from spontaneous symmetry breaking 
that also fully decouple the metric fluctuations and
the coefficient fluctuations.

For the potential terms for the two-tensor $s_\ab$, 
these are functions of the 4 possible independent scalar traces of products
of the tensor $s_{\al\be} g^{\be\ga}$ \cite{Kostelecky:2009b}.
Note that this includes the possibility of having "steep" potentials that vanish 
in the linearized or higher-order approximations.
For example, 
one useful potential is a function of $s^\al_{\pt{\al}\al}$ via
\beq
V= \fr 14 \la  \left( s^\al_{\pt{\al}\al} -x  \right)^4,
\label{pot1}
\eeq
where $x$ is a constant.
When expanding around a flat background, 
it can be shown that
the leading order effects in the field equations are of cubic order in the fluctuations.

\section{Covariant construction of Lagrange density}
\label{covariant}

As an alternative method to section \ref{general}, 
we can construct an fully covariant lagrangian for the gravitational sector 
including the $s_{\al\be}$ terms.  
We seek kinetic lagrangian terms that would produce the ones in the expansion in the previous section.
These terms were included as a second order in $s_\ab$ expansion in Ref.\ \cite{O'Neal:2021}.
These are labeled $a_1-a_{11}$ and obtained by considering all possibilities of scalars contractions of $s_\ab s_\gd R_{\mu\nu\ka\la}$ and $\nabla_\ga s_\ab \nabla_\de s_\mn$, 
and discarding surface terms:
\bea
\cL_{\rm cov} &=& \fr {\sqrt{-g}}{2\ka } \Big[ 
a_1 s^\la_{\pt{\la}\la} R
+a_2 s_\mn R^\mn
+a_3\tfrac 12 (\nabla_\mu s_{\nu\la}) (\nabla^\mu s^{\nu\la}) 
+a_4 \tfrac 12 (\nabla_\mu s^{\mu\la}) (\nabla_\la s^\be_{\pt{\be}\be}) 
\nonumber\\
&&
+a_5 \tfrac 12 (\nabla_\mu s^{\mu\la}) (\nabla_\nu s^\nu_{\pt{\be}\la} )
+a_6 \tfrac 12 ( \nabla_\mu s^\nu_{\pt{\be}\nu} ) (\nabla^\mu s^\la_{\pt{\be}\la} )
+a_7 s_\mn s_{\ka\la} R^{\mu\ka\nu\la}
+a_8 s_\mn s^{\mu}_{\pt{\mu}\la} R^{\nu\la}
\nonumber\\
&&
+a_9 s^\la_{\pt{\la}\la} s_\mn R^\mn
+a_{10} s^\mn s_\mn R
+a_{11} s^\la_{\pt{\la}\la} s^\mu_{\pt{\mu}\mu} R 
\Big].
\label{covAct}
\eea

The covariant terms can be matched to the cubic 
and quadratic terms by expanding them to the relevant orders.
The expansion, however, 
is cumbersome, 
hence the advantage of the earlier construction.
We find the following matches at quadratic order in the 
expansion:
\bea
\cL_{a_1} &=& \frac {a_1}{2\ka} \cM^{(2)}_{10} + O(3) +O(\sb^2)
\nonumber\\
\cL_{a_2} &=& \frac {a_2}{2\ka} \cM^{(2)}_{9} + O(3)+O(\sb^2)
\nonumber\\
\cL_{a_3} &=& \frac {a_3}{2\ka} 
\left( \cM^{(2)}_{5} - \cM^{(2)}_{3} \right)+ O(3)+O(\sb^2)
\nonumber\\
\cL_{a_4} &=& \frac {a_4}{4\ka} \left( \cM^{(2)}_{2} 
+ \cM^{(2)}_{7} + \frac 12 \cM^{(2)}_{8} \right) +O(3)+O(\sb^2)
\nonumber\\
\cL_{a_5} &=& \frac {a_5}{2\ka} \left( \cM^{(2)}_{5} 
+ \frac 12 \cM^{(2)}_{7} \right) +O(3)+O(\sb^2)
\nonumber\\
\cL_{a_6} &=& \frac {a_6}{2\ka} \cM^{(2)}_{8} + O(3)+O(\sb^2)
\nonumber\\
\cL_{a_7} &=& \frac {a_7}{2\ka} \left( \cM^{(2)}_{6} 
+ \frac 12 \cM^{(2)}_{7} \right) +O(3)+O(\sb^2)
\nonumber\\
\cL_{a_8} &=& \frac {a_8}{2\ka} \left( \cM^{(2)}_{3} 
+ \cM^{(2)}_{5} \right) +O(3)+O(\sb^2)
\nonumber\\
\cL_{a_9} &=& \frac {a_9}{2\ka} \left( \cM^{(2)}_{2} 
+ \cM^{(2)}_{4}-2 \cM^{(2)}_{8}
+ \frac 12 \sb^{\al}_{\pt{\al}\al} \cM^{(2)}_{9} \right) 
+O(3)+O(\sb^2)
\nonumber\\
\cL_{a_{10}} &=& \frac {a_{10}}{\ka} \cM^{(2)}_{1}  +O(3)+O(\sb^2)
\nonumber\\
\cL_{a_{11}} &=& \frac {a_{11}}{\ka} \sb^{\al}_{\pt{\al}\al}  \cM^{(2)}_{10}  +O(3)+O(\sb^2)
\label{covmatch}
\eea
Here $O(3)$ denotes terms of cubic order in fluctuations 
$h^3$, $\st^3$, $h^2 \st$, etc. while the $O(\sb^2)$ 
includes terms of second order in the vacuum values $\sb_\mn$.
Note that one could have started with the covariant expansion 
instead of going through the process of the earlier section in the work.
However, 
the quadratic and cubic expansions 
are more useful for approximate solutions 
for weak fields, 
for which one has to expand, 
order by order, in any case.
Furthermore, 
the conservation law construction can be used for more general tensor couplings, 
for which it may be challenging to construct the covariant terms.
The terms in \rf{covAct} are a subset of the recent construction
in Ref. \cite{Kostelecky:2021}.

\section{Field equations }
\label{field equations}

We turn to the field equations from the action above.
Considering the large number of terms in tables above, 
abbreviations will be used for different pieces of the field equations.
Multiple partial derivatives will be abbreviated as $\prt_{\al\be ...}=\prt_\al \prt_\be ...$.
Variation of the entire Lagrange density (through cubic order) with respect to the metric fluctuations yields the field equations in the form
\bea
\fr {\de \cL}{\de h_\mn} &=& -\frac {1}{2\ka} (G^{(1)})^\mn 
+ \frac 12 T_M^\mn + \frac 12 \ta^\mn
+ \frac {\de \cL_V}{\de h_\mn}
+ \al_1 \sb^\ab (\prt^\mn - \et^\mn \Box) \st_\ab
+ \al_2 ( \sb^{\al( \mu} \prt^{\nu )} \prt_\al -\frac 12 \eta^\mn \sb^\ab \prt_\ab ) \st
\nonumber\\
&&
+ \al_3 (- \sb^{\al (\mu} \Box \st^{\nu )}_{\pt{\nu}\al} 
+ \sb^\ab \prt_\be \prt^{(\mu } \st^{\nu )}_{\pt{\nu} \al} ) 
-\frac 12 \et^\mn \sb^\ab \prt_{\ga\be} \st_\al^{\pt{\al}\ga } ) 
+\al_5 ( \sb^{\al( \mu}  \prt^{\nu )} \prt_\be \st_\al^{\pt{\al}\be} 
-\frac 12 \et^\mn \sb^\ab \prt_{\ga\be} \st_\al^{\pt{\al}\ga} )
\nonumber\\
&&
+\al_6 ( \sb^\ab \prt_\ab -2 \sb^{\al(\mu} \prt_{\be\al} \st^{\nu )\be} ) 
+\al_7 \sb^\mn \prt_\ab \st^\ab
+ \al_8 \sb^\mn \Box \st 
\nonumber\\
&&
+ \al_9 (2 (\Gt^{(1)})^\mn + 2 \sb_\ab (\cG^{(1)})^{\al\mu\be\nu}) 
+(\al_9 +\al_{10}) [(\prt^\mn - \et^\mn \Box) (\st - \sb^\ab h_\ab ) - \sb^\mn R^{(1)}) ],
\label{metricFE}
\eea
where $\ta^\ab$ contains the full variation of the cubic terms 
$\ta^\mn = 2 \de \cL^{(3)}/\de h_\mn$, 
the factor of $2$ chosen so that it matches the GR pseudo tensor in the correct limit.
Here $\cL_V$ is the portion of the Lagrange density with the potential energy 
function for $\st_\mn$, 
and $T_M^\mn$ comes from the variation of the matter sector with respect
to the metric fluctuations.
Similarly, 
the variation of the Lagrange density with respect to the two-tensor fluctuations
$\st_\ab$ is 
\bea
\fr {\de \cL}{\de \st_\mn} &=& (\ta_s)^\mn + \frac {\de \cL_V}{\de \st_\mn}+
\al_1 \sb^\mn R^{(1)}
+\al_2 
[ 2 (\Gt^{(1)})^\mn -\frac 12 \et^\mn 
( \Box \st + \sb^\ab (\prt_\ab h - 2 \prt_{\ga\be} h_\al^{\pt{\al}\ga}))] 
\nonumber\\
&&
+ \al_3 [\frac 12 \Box \st^\ab -\frac 12 \sb^{\al(\mu} \prt^{\nu )} \prt_\al h 
+ \sb^{\al (\mu } \prt_\ab h^{\nu) \be} - \sb^{\al(\mu } \Box h^{\nu)}_{\pt{\nu}\al}] 
- 2 \al_4 (\Gt^{(1)})^\mn 
\nonumber\\
&&
+ \al_5 [-\frac 12 \Box \st^\ab - \frac 12 \sb^{\al(\mu} \prt^{\nu )} \prt_\al h 
+ \sb^{\al (\mu } \prt^{\nu )} \prt_\be h_\al^{\pt{\al}\be}]
\nonumber\\
&&
+\al_6 [ -\Box \st^\mn +2 \prt_\al \prt^{(\mu} \st^{\nu)\al} 
  +\sb^\ab ( \prt_\ab h^\mn - \prt_\be \prt^{(\mu} h^{\nu)}_{\pt{\nu}\al}) ]
\nonumber\\
&&
+\al_7 ( \Box \st^\mn - 2 \prt_\al \prt^{(\mu} \st^{\nu)\al} +\sb^\ab \prt^\mn h_\ab)
- \al_8 \et^\mn \Box (\st - \sb^\ab h_\ab )
\nonumber\\
&&
+ 2 \al_9 (R^{(1)})^\mn 
+\al_{10} \et^\mn R^{(1)},
\label{sFE}
\eea
where $(\ta_s)^\mn = \de \cL^{(3)}/\de \st_\mn $.
The matter sector will have its own field equations which depend on the system
to be modeled. 
We note here that when the field equations hold, 
the stress-energy tensor for matter will be conserved, 
a fact that can be exploited in solving the post-Newtonian limit.

The generic case of equations
\rf{metricFE} and \rf{sFE} with arbitrary parameters 
contains ``ghost" kinetic terms for the fluctuations $\st_\ab$, 
when treated as a field theory on a flat background.
For example, a kinetic term of the form $\cL \sim \st^\ab \Box \st_\ab$
will contain non-positive definite contributions to the Hamiltonian density, 
and could thereby result in unstable solutions.
It is of general interest to study this issue in detail, 
but we do not attempt it here.

We note that this effective field theory construction represents 
a test framework for Lorentz and diffeomorphism breaking.
Rather than just the $9$ coefficients in \rf{lag2}, 
we now have up to $10$ arbitrary parameters $\al_1- \al_{10}$ and $9$ coefficients $\sb_\mn$, 
for a total of $19$ quantities describing Lorentz and diffeomorphism violation
for gravity.

\section{Summary and Outlook}
\label{summary and outlook}

In this work we described the method to construct quadratic and cubic effective actions for weak-field gravity studies of spontaneous symmetry breaking.
The main results are the Lagrange density terms in equations \rf{Lform} and 
\rf{Lform3} and the Tables \ref{M2} and \ref{M3}.
Furthermore, 
we showed that complete decoupling of the fluctuations
from the metric fluctuations cannot occur at the level of the action
with only second derivatives when going to cubic order in the fluctuations.

The field equations for the metric fluctuations and the two-tensor fluctuations
are given in equations \rf{sFE} and \rf{metricFE}.
Future work will solve these equations in the post-Newtonian limit, 
with near-field and radiation zone solutions for applications in astrophysics.
Some preliminary analysis of the field equations in \rf{sFE} and \rf{metricFE}
shows that the near-field post-Newtonian metric contains many of the same
fluid potential function terms that were found in Ref.\ \cite{Bailey:2006}.
We conjecture that including all of the extra terms in the Lagrange density
could potentially provide an interesting re-analysis of the match
to the Parametrized Post-Newtonian metric \cite{Will:1971, Will:1973}.
The match of the action above to vector models of Lorentz violation 
should also be revisited
\cite{Kostelecky:1989,Jacobson:2001,Seifert:2009}.
Finally, 
the method adopted in this work for the two-tensor $s_\mn$ 
could be continued and applied to other gravity sector coefficients, 
such at the $t_\abgd$ coefficients, 
or non-minimal coefficients.

Analysis could also proceed with the general covariant action \rf{covAct}, 
such as exploring black hole solutions or cosmology solutions.
Progress has already been made studying strong-field gravity solutions 
with vector and tensor models of spontaneous Lorentz-symmetry breaking 
\cite{Casana:2018,Maluf:2021}.
Complimentary analysis with \rf{covAct} would be of interest.

\section{Acknowledgements}

This work was supported in part by the National Science Foundation 
under Grant No. 1806871.
K.\ O'Neal-Ault and M.\ Mewes provided valuable comments on the manuscript.

\reftitle{References}



\begin{thebibliography}{99}

\bibitem{Hehl:1976}
  F.~W.~Hehl \etal, "General relativity with spin and torsion: Foundations and prospects," Rev.\ Mod.\ Phys.\ {\bf 48}, 393 (1976).
\bibitem{Kostelecky:1988}  
  V.~A.~Kosteleck\'y and S.~Samuel,
  ``Spontaneous Breaking of Lorentz Symmetry in String Theory,''
  Phys.\ Rev.\ D {\bf 39}, 683 (1989).
\bibitem{Hees:2016} 
  A.~Hees \etal,
  ``Tests of Lorentz symmetry in the gravitational sector," 
  Universe {\bf 2}, 4 (2016).
\bibitem{Kostelecky:2008} 
  V.~A.~Kosteleck\'y and N.~Russell,
  ``Data Tables for Lorentz and CPT Violation,''
  Rev.\ Mod.\ Phys.\  {\bf 83}, 11 (2011), arXiv:0801.0287.
\bibitem{Tasson:2016}
  J.~Tasson, ``The Standard-Model Extension and Gravitational Tests,"
  Symmetry {\bf 8}, 111 (2016). 
\bibitem{Will:2014}
  C.M.~Will, ``The Confrontation between General Relativity and Experiment,"
  Living Rev.\ Relativity {\bf 17}, 4 (2014).
\bibitem{Yunes:2016}
  N.\ Yunes, K.\ Yagi, and F.\ Pretorius, 
  ``Theoretical physics implications of the binary black-hole mergers GW150914 and GW151226," Phys.\ Rev.\ D {\bf 94}, 084002.
\bibitem{Kostelecky:1995} 
  V.~A.~Kosteleck\'y and R.~Potting,
  ``CPT, strings, and meson factories,''
  Phys.\ Rev.\ D {\bf 51}, 3923 (1995). 
\bibitem{Colladay:1997} 
  D.~Colladay and V.~A.~Kosteleck\'y,
  ``CPT violation and the standard model,''
  Phys.\ Rev.\ D {\bf 55}, 6760 (1997). 
\bibitem{Colladay:1998} 
  D.~Colladay and V.~A.~Kosteleck\'y,
  ``Lorentz violating extension of the standard model,''
  Phys.\ Rev.\ D {\bf 58}, 116002 (1998). 
\bibitem{Kostelecky:2004} 
  V.~A.~Kosteleck\'y,
  ``Gravity, Lorentz violation, and the standard model,''
  Phys.\ Rev.\ D {\bf 69}, 105009 (2004).
\bibitem{LePoncinLafitte:2016}  
  C. Le Poncin-Lafitte, A. Hees, and S. Lambert, ``Lorentz symmetry and Very Long Baseline Interferometry",  
  Phys.\ Rev.\ D {\bf 94}, 125030 (2016).
\bibitem{Abbott:2017}
  B.~P.~Abbott, \etal, 
  "Gravitational Waves and Gamma-Rays from a Binary Neutron Star Merger: GW170817 and GRB 170817A,"
  Astrophys.\ J.\ {\bf 848} (2017) L13.
\bibitem{Bailey:2009}
  Q.G.~Bailey, ``Time-delay and Doppler tests of the Lorentz symmetry of gravity,"
  Phys.\ Rev.\ D {\bf 80}, 044004 (2009).
\bibitem{Bailey:2006}  
  Q.~G.~Bailey and  V.~A.~Kosteleck\'y,
  ``Signals for Lorentz Violation in Post-Newtonian Gravity,''
  Phys.\ Rev.\  D {\bf 74}, 045001 (2006).
\bibitem{Kostelecky:2015} 
  V.~A.~Kosteleck\'y and J.~Tasson,
  ``Constraints on Lorentz violation from gravitational Cherenkov radiation,''
  Phys.\ Lett.\ B {\bf 749}, 551 (2015)
\bibitem{Kostelecky:2016} 
  V.~A.~Kosteleck\'y and M.~Mewes,
  ``Testing local Lorentz invariance with gravitational waves,''
  Phys.\ Lett.\ B {\bf 757}, 510 (2016).
\bibitem{Kostelecky:2009a}
  V.~A.~Kosteleck\'y and J.~ Tasson,``Prospects for Large Relativity Violations in Matter-Gravity Couplings,"
  Phys.\ Rev.\ Lett.\ {\bf 102}, 010402 (2009).
\bibitem{Kostelecky:2011}
  V.~A.~Kosteleck\'y and J.~Tasson, ``Matter-gravity couplings and Lorentz violation,"
   Phys.\ Rev.\ D {\bf 83}, 016013 (2011).
\bibitem{Bourgoin:2016}
  A.~Bourgoin \etal,
 ``Testing Lorentz symmetry with Lunar Laser Ranging," 
 Phys.\ Rev.\ Lett.\ {\bf 117}, 241301 (2016).
\bibitem{Mueller:2008}
  H.~ Mueller,
  ``Atom interferometry tests of the isotropy of post-Newtonian gravity," 
  Phys.\ Rev.\ Lett.\ {\bf 100}, 031101 (2008).
\bibitem{Flowers:2017}
  N.~A.~Flowers, C.~Goodge, and J.~D.~Tasson, 
  ``Superconducting-Gravimeter Tests of Local Lorentz Invariance,"
  Phys.\ Rev.\ Lett.\ {\bf 119}, 201101 (2017).
\bibitem{Shao:2018}
  C.~G.~Shao \etal,
  ``Limits on Lorentz violation in gravity from worldwide superconducting gravimeters,"
  Phys.\ Rev.\  D {\bf 97}, 024019 (2018).
\bibitem{Hees:2015}
  A.~Hees \etal, ``Testing Lorentz symmetry with planetary orbital dynamics,"
  Phys.\ Rev.\ D {\bf 92}, 064049 (2015).
\bibitem{LShao:2014}
  L.~Shao, ``Tests of Local Lorentz Invariance Violation of Gravity in the Standard Model Extension with Pulsars,"
  Phys.\ Rev.\ Lett.\ {\bf 112}, 111103 (2014).
\bibitem{Bailey:2016} 
  Q.~G.~Bailey, 
  ``Anisotropic cubic curvature couplings,''
  Phys.\ Rev.\ D {\bf 94}, 065029 (2016)
\bibitem{Bonder:2015}
  Y.~Bonder, ``Lorentz violation in the gravity sector: The t puzzle,"
  Phys.\ Rev.\ D {\bf 91}, 125002 (2015).
\bibitem{Bonder:2017}
  Y.~Bonder and G.~Leon, ``Inflation as an amplifier,"
  Phys.\ Rev.\ D {\bf 96}, 044036 (2017).
\bibitem{Bailey:2019}
  Q.G.~Bailey, ``Recent Developments in Spacetime-Symmetry tests in Gravity,"
  in CPT and Lorentz Symmetry VIII, edited by R.~Lehnert (World Scientific, Singapore, 2020).
\bibitem{Bonder:2020}
  Y.~Bonder and C.~Peterson, 
  ``Explicit Lorentz violation in a static and spherically-symmetric spacetime,"
\bibitem{O'Neal:2021}
   K.~O'Neal-Ault, Q.~G.~Bailey, and N.~A.~Nilsson, ``3+1 formulation of the standard model extension,"
   Phys.\ Rev.\ D {\bf 103}, 044010 (2021).
\bibitem{Xu:2020}
  R.~Xu, J.~Zhao, and L.~Shao, ``Neutron star structure in the minimal gravitational Standard-Model Extension and the implication to continuous gravitational waves,"
  Phys.\ Lett.\ B {\bf 803}, 135283 (2020).
\bibitem{Xu:2021}
   R.\ Xu, Y.\ Gao, and L.\ Shao, ``Precession of spheroids under Lorentz violation and observational consequences for neutron stars,"
   arXiv:2012.01320.
\bibitem{Bonder:2021}
  Y.~Bonder and C.\ Peterson, 
  ``Spontaneous Lorentz violation and asymptotic flatness,"
  arXiv:2103.07611.
\bibitem{Kostelecky:2018} 
  V.~A.~Kosteleck\'y and M.~Mewes,
  ``Lorentz and Diffeomorphism Violations in Linearized Gravity,''
  Phys.\ Lett.\ B {\bf 779}, 136 (2018)
\bibitem{Seifert:2009}
  M.~Seifert, ``Vector models of gravitational Lorentz symmetry breaking,"
  Phys.\ Rev.\ D {\bf 79}, 124012 (2009).
\bibitem{Altschul:2010}
  B.~Altschul, Q.G.~Bailey, and V.~A.~Kosteleck\'y, "Lorentz violation with an antisymmetric tensor,"
  Phys.\ Rev.\ D {\bf 81}, 065028 (2010).
\bibitem{Seifert:2018}
  M.~Seifert, "Lorentz-Violating Gravity Models and the Linearized Limit," 
  Symmetry {\bf 10}, 490 (2018).
\bibitem{Kostelecky:2021}
   V.~A.~Kosteleck\'y and Z.Li, ``Backgrounds in gravitational effective field theory,"
   Phys.\ Rev.\ D {\bf 103}, 024059 (2021).
\bibitem{Bailey:2018}
   Q.G.~Bailey and C.D.~Lane, ``Relating Noncommutative SO(2,3)* Gravity to the Lorentz-Violating Standard-Model Extension,"
   Symmetry {\bf 10}, 480 (2018).
\bibitem{Bailey:2015} 
  Q.~G.~Bailey, V.~A.~Kosteleck\'y, and R.~Xu,
  ``Short-range gravity and Lorentz violation,''
  Phys.\ Rev.\ D {\bf 91}, 022006 (2015).
\bibitem{Bluhm:2005}
  R.~Bluhm and V.~A.~Kosteleck\'y,
  ``Spontaneous Lorentz Violation, Nambu-Goldstone Modes, and Gravity,"
  Phys.\ Rev.\ D {\bf 71}, 065008 (2005).
\bibitem{Bluhm:2008}
  R.~Bluhm, S.H.~Fung, and V.~A.~Kosteleck\'y, 
  ``Spontaneous Lorentz and diffeomorphism violation, massive modes, and gravity,''
  Phys.\ Rev.\ D  {\bf 77}, 065020 (2008). 
\bibitem{Bluhm:2015}
  R.~Bluhm, 
  ``Explicit versus spontaneous diffeomorphism breaking in gravity,''
  Phys.\ Rev.\ D  {\bf 91}, 065034 (2015). 
\bibitem{Bluhm:2016}
  R.~Bluhm and A.~Sehic, 
  ``Noether identities in gravity theories with nondynamical backgrounds
   and explicit spacetime symmetry breaking,''
   Phys.\ Rev.\ D  {\bf 94}, 104034 (2016). 
\bibitem{Bluhm:2019}
   R.~Bluhm, H.~Bossi, and Y.~Wen, 
  ``Gravity with explicit spacetime symmetry breaking and the
   standard model extension,"
  Phys.\ Rev.\ D  {\bf 100}, 084022 (2019). 
\bibitem{Bluhm:2021}
   R.~Bluhm and Y.~Yang, ``Gravity with Explicit Diffeomorphism Breaking," Symmetry {\bf 13}, 660 (2021).
\bibitem{xTras:2014}
  T.~Nutma, "xTras: a field-theory inspired xAct package for Mathematica," 
  Comput.\ Phys.\ Commun.\ {\bf 185}, 1719 (2014). doi:10.1016/j.cpc.2014.02.006.
\bibitem{Kostelecky:2009b}
   V.~A.~Kosteleck\'y and R.~Potting, ``Gravity from spontaneous Lorentz violation,"
   Phys.\ Rev.\ D {\bf 79}, 065018 (2009).
\bibitem{Will:1971}
   C.M.~Will, ``Theoretical Frameworks for Testing Relativistic Gravity. II. Parametrized Post-Newtonian Hydrodynamics, and the Nordtvedt Effect,"
   Astrophys.\ J.\ {\bf 163}, 611 (1971).
\bibitem{Will:1973}
   C.M.~Will, ``Relativistic Gravity tn the Solar System. 111. Experimental Disproof of a Class of Linear Theories of Gravitation,"
   Astrophys.\ J.\ {\bf 185}, 31 (1973).
\bibitem{Kostelecky:1989}
   V.~A.~Kosteleck\'y and S.\ Samuel, 
   ``Gravitational phenomenology in higher-dimensional theories and strings,"
   Phys.\ Rev.\ D {\bf 40}, 1886 (1989).
\bibitem{Jacobson:2001}
   T.\ Jacobson and D.\ Mattingly, ``Gravity with a dynamical preferred frame,"
   Phys.\ Rev.\ D {\bf 64}, 024028 (2001).
\bibitem{Casana:2018}   
   R.\ Casana \etal, "Exact Schwarzschild-like solution in a bumblebee gravity model,"
   Phys.\ Rev.\ D {\bf 97}, 104001 (2018). 
\bibitem{Maluf:2021}
   R.V.\ Maluf and J.C.S.\ Neves, 
   ``Black holes with a cosmological constant in bumblebee gravity,"
   Phys.\ Rev.\ D {\bf 103}, 044002 (2021).  
\end{thebibliography}
\end{document}